\documentclass[11pt]{article}
\usepackage{latexsym}
\usepackage{amsfonts}
\usepackage{amsmath}
\usepackage{amsthm}
\usepackage{graphicx}
\usepackage{epic}
\def\a{\alpha}
\def\b{\beta}
\def\g{\gamma}
\def\d{\delta}

\def\l{\lambda}

\def\m{\mu}
\def\n{\nu}
\def\r{\rho}

\def\p{\pi}

\def\cA{{\mathcal A}}

\def\be{\begin{equation}}
\def\ee{\end{equation}}
\def\beq{\begin{eqnarray}}
\def\eeq{\end{eqnarray}}
\def\nn{\nonumber}

\def\ce{{\mathcal E}}
\def\cf{{\mathcal F}}

\def\ch{{\mathcal H}}

\def\ck{{\mathcal K}}

\def\cm{{\mathcal M}}
\def\cn{{\mathcal N}}

\def\cv{{\mathcal V}}

\def\RR{{\mathbb{R}}}

\def\ZZ{{\mathbb{Z}}}

\newcommand{\ft}[2]{{\textstyle {\frac{#1}{#2}} }}

\newcommand{\E}{E_{10}}

\newcommand{\Ref}[1]{(\ref{#1})}
\newcommand{\non}{\nonumber\\}

\newcommand{\bqn}{\begin{eqnarray}}\newcommand{\eqn}{\end{eqnarray}}

\title{Cosmological Singularities,
Billiards and Lorentzian Kac-Moody Algebras}
\author{Thibault Damour}
\date{\it Institut des Hautes \'Etudes Scientifiques, 35 route de Chartres,
91440 Bures-sur-Yvette, France}

\begin{document}

\maketitle

\begin{abstract}
The structure of  the general, inhomogeneous solution of (bosonic) Einstein-matter
systems in the vicinity of a cosmological
singularity is considered. We review the proof (based on ideas of Belinskii-Khalatnikov-Lifshitz and
technically  simplified by the use of the Arnowitt-Deser-Misner Hamiltonian
formalism) that the asymptotic behaviour, as one approaches the singularity,
of the general solution is  describable, at each (generic)
spatial point, as a billiard motion in an auxiliary Lorentzian space.
For certain Einstein-matter systems, notably for pure Einstein gravity in
any spacetime dimension $D$ and for the particular Einstein-matter systems
arising in String theory, the billiard tables describing asymptotic
cosmological behaviour are found to be identical to the Weyl chambers of
some Lorentzian Kac-Moody algebras. In the case of the bosonic sector of
supergravity in 11 dimensional spacetime the underlying Lorentzian algebra
is that of the hyperbolic Kac-Moody group $E_{10}$, and there exists some
evidence of a correspondence between the general solution of the
Einstein-three-form system and a null geodesic in the infinite dimensional
coset space $E_{10} / K (E_{10})$, where $K (E_{10})$ is the maximal
compact subgroup of $E_{10}$.

\end{abstract}

{\sl It is a pleasure to dedicate this review to Stanley Deser,
a dear friend and a great physicist to whom I owe a lot.}

\section{Introduction and overview}

A remarkable connection between the asymptotic behavior
of certain Einstein-matter systems near a cosmological singularity  and
billiard motions in the Weyl chambers of some corresponding Lorentzian
Kac--Moody algebras was uncovered in a series of
works \cite{DH1,DH2,DH3,DHJN,DdBHS,DHN2,Damour:2002et}. This simultaneous appearance of {\it billiards}
(with {\it chaotic} properties in important physical cases) and of an underlying
{\it symmetry} structure (infinite-dimensional Lie algebra) is an interesting
fact, which deserves to be studied in depth. Before explaining  the
techniques (notably the Arnowitt-Deser-Misner Hamiltonian formalism \cite{ADM})
that have been used to uncover this fact, we will start by reviewing
previous related works, and by stating the main results of this billiard/symmetry connection.

The simplest example of this connection concerns the
pure Einstein system in $D=3+1$-dimensional space-time.
The Einstein equations ($R_{\m\n}(g_{\a\b}) = 0)$ are non-linear PDE's
for the metric components. Near a cosmological spacelike singularity, here chosen as $t=0$, the spatial gradients are expected
to become negligible compared to time derivatives (${\partial \over
\partial x^i } << {\partial \over
\partial t}$); this then suggests the decoupling of spatial points and allows for an approximate
treatment in which one replaces the  above partial differential equations by (a 3-dimensional family of)
ordinary differential equations. Within this simplified context,  Belinskii,
Khalatnikov and Lifshitz (BKL)  gave a description \cite{BKL1,BKL2,BKL3} of
the asymptotic behavior of the general
solution of  Einstein's equations, close to the singularity,  and showed that it can be described as a chaotic \cite{KLL,Bar}
sequence of generalized Kasner solutions. The Kasner metric is of the type
\beq g_{\a\b}(t) dx^{\a}dx^{\b} = - N^2 dt^2 + A_1 t^{2p_1} dx_1^2
+A_2 t^{2p_2} dx_2^2 +A_3 t^{2p_3} dx_3^2  \eeq where the
constants $p_i$ obey\footnote{In the $N=1$ gauge, they also obey $p_1+p_2+p_3=1$.} \beq \label{aux_metric}
\overrightarrow{p}^2 = p_1^2+p_2^2+p_3^2- (p_1+p_2+p_3)^2= 0. \eeq 
An exact Kasner solution, with a given set of $A_i$'s and $p_i$'s, can
be represented by a null line in a 3-dimensional auxiliary Lorentz space with
coordinates $p_1, p_2, p_3$ equipped with the metric given by the quadratic form
 $\overrightarrow{p}^2$ above.
The auxiliary Lorentz space can be radially projected on the unit hyperboloid or further on the Poincar\'e disk (i.e.
on the hyperbolic plane $H_2$):  the projection of a null line is a geodesic on the hyperbolic plane.

\begin{figure}[ht]
\centerline{\includegraphics[scale=.8]{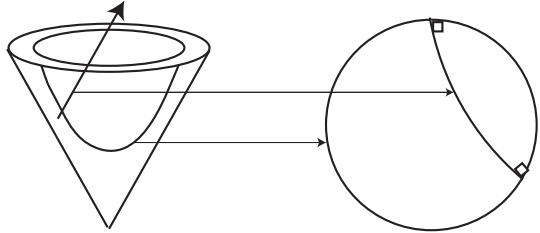}}
\caption{\small{Lorentz space and projection on Poincar\'e disk.}}
\end{figure}

BKL showed  that, because of non-linearities in  Einstein's equations,  the generic solution behaves as a succession of Kasner epochs, {i.e.},  to a broken null line in the auxiliary Lorentz space, or a broken geodesic on the Poincar\'e disk.
This broken geodesic motion is a ``billiard motion'' (seen either in Lorentzian space or in hyperbolic space).

\begin{figure}[ht]
\centerline{\includegraphics[scale=.8]{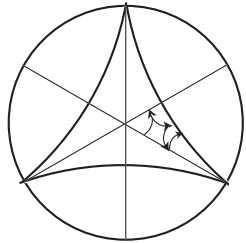}}
\caption{\small{Picture of chaotic cosmological behavior for $3+1$ gravity}}
\end{figure}

The billiard picture naturally follows from the Hamiltonian approach to cosmological behavior and was first obtained in the homogeneous (Bianchi IX) four-dimensional case \cite {Chitre,Misnerb} and then extended to higher space-time dimensions with $p$-forms and dilatons \cite{DH3,DHN2,Damour:2002et,Kirillov1993, KiMe,IvKiMe94,IvMe}.
Recent work \cite{Damour:2002et}
has improved the derivation of the billiard picture by using the Iwasawa decomposition of the spatial metric.
Combining this decomposition with the Arnowitt-Deser-Misner
Hamiltonian formalism highlights the mechanism by which all variables except the scale factors and the dilatons get asymptotically frozen.
The non-frozen variables (logarithms of scale factors and dilatons) then undergo a billiard motion.
This billiard motion can be seen either in Lorentzian space or, after radial projection, on hyperbolic
space (see below for details).

A remarkable connection was also established \cite{DH1,DH2,DH3,DHJN,DdBHS,DHN2,Damour:2002et} between certain specific Einstein-matter systems and Lorentzian Kac-Moody (KM) algebras \cite{Kac}.  In the leading asymptotic approximation, this connection is simply that the Lorentzian billiard table within which the motion is confined can be identified with the  Weyl chamber of some corresponding Lorentzian KM algebra. This can happen only when many conditions are met: in particular, (i) the billiard table must be a Coxeter polyhedron (the dihedral angles between adacent walls must be integer submultiples of $\pi$) and ii) the billiard must be a simplex. Surprisingly, this occurs in many physically interesting Einstein-matter systems. For instance, pure Einstein gravity in $D$ dimensional space-time corresponds to the
Lorentzian KM algebra $AE_{D-1}$ \cite{DHJN} which is the overextension of the finite Lie algebra $A_{D-3}$:  for $D=4$, the algebra is $AE_3$ the Cartan
matrix of which is given by
\beq A= \left( \begin{array}{rrr} 2 & -1 & 0 \\
-1 & 2 &-2 \\  0 & -2& 2 \\  \end{array}  \right) \eeq Chaotic billiard tables have finite volume in hyperbolic space, while non-chaotic ones have infinite volume; as a consequence, chaotic billiards are associated with {\it hyperbolic} KM algebras; this happens to be the case for pure gravity when $D\leq 10$.

Another connection between physically interesting Einstein-matter systems and
KM algebras concerns the low-energy bosonic effective actions
arising in string and $M$ theories. Bosonic string theory in any space-time
dimension $D$ is related to the Lorentzian KM algebra $DE_D$
\cite{DH3,DdBHS}. The latter algebra is the canonical Lorentzian extension
of the finite-dimensional algebra $D_{D-2}$. The various superstring
theories (in the critical dimension $D=10$) and $M$-theory have been found
\cite{DH3} to be related either to $E_{10}$ (when there are two
supersymmetries in $D=10$, i.e. for type IIA, type IIB and $M$-theory) or
to $BE_{10}$ (when there is only one supersymmetry in $D=10$,
i.e. for type I and II heterotic theories), see the table below.
A construction of the Einstein-matter systems related  to the canonical Lorentzian extensions of
{\it all}  finite-dimensional Lie algebras $A_n$, $B_n$, $C_n$, $D_n$,
$G_2$, $F_4$, $E_6$, $E_7$ and $E_8$ (in
the above ``billiard'' sense) is presented in Ref.~\cite{DdBHS}.
See also Ref.~\cite{dBS} for the identification
of all hyperbolic KM algebras whose Weyl chambers are  Einstein billiards.

The correspondence between the
specific Einstein--three-form system (including a Chern--Simons term)
describing the bosonic sector of 11-dimensional supergravity (also known as
the ``low-energy limit of $M$-theory'') and the hyperbolic KM group
$E_{10}$ was studied in more detail in \cite{DHN2}. Reference  \cite{DHN2} introduces a formal expansion of the field
equations in terms of positive roots, i.e. combinations $\alpha =
\Sigma_i \, n^i \, \alpha_i$ of simple roots of $E_{10}$, $\alpha_i$, $i =
1,\ldots , 10$, where the $n^i$'s are integers $\geq 0$. It is then useful to
{\it order} this expansion according to the {\it height} of the positive root
$\alpha = \Sigma_i \, n^i \,
\alpha_i$, defined as ${\rm ht} (\alpha) = \Sigma_i \, n^i$. The correspondence
discussed above between the {\it leading} asymptotic evolution near a
cosmological singularity (described by a billiard) and Weyl chambers of
KM algebras involves only the terms in the field equation whose
height is ${\rm ht} (\alpha) \leq 1$. By contrast, the authors of Ref. \cite{DHN2} managed to show,
by explicit calculation, that there exists a way to define, at each
spatial point $x$, a correspondence between the field variables $g_{\mu\nu}
(t,x)$, $A_{\mu\nu\lambda} (t,x)$ (and their gradients), and a (finite)
subset of the parameters defining an element of the (infinite-dimensional)
coset space $E_{10} / K(E_{10})$ where $K(E_{10})$ denotes the maximal
compact subgroup of $E_{10}$, such that the (PDE) field equations of
supergravity get mapped onto the (ODE) equations describing a null geodesic
in $E_{10} / K(E_{10})$ {\it up to terms of height} 30. This tantalizing
result suggests that the infinite-dimensional hyperbolic Kac--Moody group $E_{10}$
may be a ``hidden symmetry'' of supergravity in the sense of mapping
solutions onto solutions (the idea that $E_{10}$ might be
a symmetry of supergravity was first suggested by
Julia long ago \cite{Julia,Julia2}). Note that the conjecture here is that the { \it continuous} group
$E_{10}(\RR)$ be a hidden symmetry group of { \it classical} supergravity.
At the {\it quantum} level, i.e. for M theory, one expects only a discrete version
of $E_{10}$, say $E_{10}(\ZZ)$, to be a quantum symmetry. See \cite{BGH}
for recent work on
extending the identification of \cite{DHN2} between roots of $E_{10}$ and symmetries
of supergravity/M-theory beyond height 30, and for references about previous suggestions
of a possible role for $E_{10}$. For earlier appearances of the Weyl groups
of the $E$ series in the context of $U$-duality see \cite{LPS,OPR,BFM}.
A series of recent papers  \cite{West,SWest,SWest2,Englert1,Englert2}
 has also explored the possible role of $E_{11}$
(a nonhyperbolic extension of $E_{10}$) as a hidden symmetry of M theory.

It is also tempting to assume that the
KM groups underlying the other (special) Einstein-matter systems
discussed above might be hidden (solution-generating) symmetries. For
instance, in the case of pure Einstein gravity in $D=4$ space-time, the
conjecture is that $AE_3$ be such a symmetry of Einstein gravity. This
case, and the correspondence between the field variables and the coset ones
is further discussed in \cite{Damour:2002et}. 

Rigorous mathematical proofs \cite{AR, Rendall:2001nx, DHRW} are however only available
for `non chaotic' billiards.

In the remainder of this paper, we will outline various arguments explaining the above results;   a more complete derivation can be found in \cite{Damour:2002et}.

\section{General Models}
The general systems considered here are of the following form
\beq &&S[{\rm g}_{MN}, \phi, A^{(p)}] = \int d^D x \, \sqrt{- {\rm
g}} \;
\Bigg[R ({\rm g}) - \partial_M \phi \partial^M \phi \nonumber \\
&& \hspace{2.5cm} - \frac{1}{2} \sum_p \frac{1}{(p+1)!} e^{\l_p
\phi} F^{(p)}_{M_1 \cdots M_{p+1}} F^{(p)  \, M_1 \cdots M_{p+1}}
\Bigg] + \dots .~~~~~~~ \label{keyaction} \eeq Units are
chosen such that $16 \pi G_N = 1$,  $G_N$ is Newton's
constant and the space-time dimension $D \equiv d+1$ is left
unspecified. Besides the standard Einstein--Hilbert term the above
Lagrangian contains a dilaton\footnote{The generalization to any number of dilatons is
straightforward.}  field $\phi$ and a number of
$p$-form fields $A^{(p)}_{M_1 \cdots M_p}$ (for $p\geq 0$).  The $p$-form field strengths $F^{(p)} = dA^{(p)}$ are
normalized as \be F^{(p)}_{M_1 \cdots M_{p+1}} = (p+1)
\partial_{[M_1} A^{(p)}_{M_2 \cdots M_{p+1}]} \equiv
\partial_{M_1} A^{(p)}_{M_2 \cdots M_{p+1}} \pm p \hbox{
permutations }. \ee As a
convenient common formulation we adopt the Einstein conformal
frame and normalize the kinetic term of the dilaton $\phi$ with
weight one with respect to the Ricci scalar. The Einstein metric
${\rm g}_{MN}$ has Lorentz signature $(- + \cdots +)$ and is used
to lower or raise the indices; its determinant is denoted by ${\rm
g}$. The dots in the action (\ref{keyaction}) above
indicate possible modifications of the field strength by
additional Yang--Mills or Chapline--Manton-type couplings
\cite{pvnetal,CM}. The real parameter $\l_p$ measures the strength
of the coupling of $A^{(p)}$ to the dilaton. When $p=0$, we assume
that $\l_0\neq 0$ so that there is only one dilaton. 

\section{Dynamics in the vicinity of a spacelike singularity}

The main technical points that will be reviewed here are the following
\begin{itemize}
\item near the singularity, $t \rightarrow 0$, due to the decoupling of space points, the Einstein's (PDE) equations
 become ODE's with respect to time.

\item The study of these ODE's near $t\to 0$, shows that the $d$ diagonal spatial metric components "$g_{ii}$" and the dilaton $\phi$ move on a billiard in an auxiliary $d+1$ dimensional Lorentz space.

\item All the
other field variables ($g_{ij}, i\neq j, A_{i_1...i_p}, \pi^{i_1...i_p})$ freeze as $t \rightarrow 0$. 

\item In many interesting cases, the billiard tables
can be identified with the fundamental Weyl chamber of an hyperbolic KM algebra.

\item For SUGRA$_{11}$, the KM algebra is $E_{10}$.  Moreover,
 the PDE's are equivalent  to
 the equations  of a null geodesic on the
coset space $E_{10} / K(E_{10}) $, up to height 30.
\end{itemize}

\subsection{Arnowitt-Deser-Misner Hamiltonian formalism}

To focus on the features relevant to the billiard picture, we
assume here that there are no Chern--Simons and no Chapline--Manton terms and that the
curvatures $F^{(p)}$ are abelian, $F^{(p)}
= d A^{(p)}$. That such additional terms do not alter the analysis has been proven in \cite{Damour:2002et}. In any pseudo-Gaussian gauge and in the temporal gauge ($g_{0i}=0$ and $A_{0 i_2...i_p}=0$, $\forall p$), the
Arnowitt-Deser-Misner Hamiltonian action \cite{ADM}
reads \beq && S\left[ g_{ij}, \pi^{ij}, \phi,
\pi_\phi, A^{(p)}_{j_1 \cdots j_p},
\pi_{(p)}^{j_1 \cdots j_p}\right] = \nonumber \\
&& \hspace{1cm} \int dx^0 \int d^d x \left( \pi^{ij} \dot{g_{ij}}
+ \pi_\phi \dot{\phi} + \frac{1}{p!}\sum_p \pi_{(p)}^{j_1 \cdots
j_p} \dot{A}^{(p)}_{j_1 \cdots j_p} - H \right)\,,
\label{GaussAction} \eeq  where the Hamiltonian density $H$ is
\beq\label{Ham}
H &\equiv&  \tilde{N} \ch \, ,\\[2mm]
\label{Ham1}\ch &=& \ck + \cm \, ,\\[2mm]
\ck &=& \pi^{ij}\pi_{ij} - \frac{1}{d-1} \pi^i_{\;i} \pi^j_{\;j}
+ \frac{1}{4} \pi_\phi^2 
+ \sum_p \frac{e^{- \lambda_p \phi}} {2 \, p!} \, \pi_{(p)}^{j_1
\cdots j_p}
\pi_{(p) \, j_1 \cdots j_p} \, ,~~~~~\\[2mm]
\cm &=& - g R + g g^{ij} \partial_i \phi \partial_j \phi + \sum_p
\frac{e^{ \lambda_p \phi}}{2 \; (p+1)!} \, g \, F^{(p)}_{j_1
\cdots j_{p+1}} F^{(p) \, j_1 \cdots j_{p+1}}\,, \eeq

\noindent and $R$ is the spatial curvature scalar. $\tilde{N} = N/\sqrt{g^{(d)}}$ is the rescaled lapse. The dynamical
equations of motion are obtained by varying the above action with
respect to the spatial metric components, the dilaton, the spatial
$p$-form components and their conjugate momenta. In addition,
there are constraints on the dynamical variables,

\beq
\ch &\approx& 0  \; \; \; \; \; \; \hbox{(``Hamiltonian constraint")}, \\[2mm]
\ch_i &\approx& 0  \; \; \; \; \; \; \hbox{(``momentum constraint")}, \\[2mm]
\varphi_{(p)}^{j_1 \cdots j_{p-1}} &\approx& 0 \; \; \; \;\; \;
\hbox{(``Gauss law" for each $p$-form), } \label{Gauss} \eeq with
\beq \ch_i &=& -2 {\pi^j}_{i|j} + \pi_\phi
\partial_i \phi + \sum_p \frac1{p!} \
\pi_{(p)}^{j_1 \cdots j_p} F^{(p)}_{i j_1 \cdots j_{p}} \,,\\[2mm]
\varphi_{(p)}^{j_1 \cdots j_{p-1}} &=& {\pi_{(p)}^{j_1 \cdots
j_{p-1} j_p}}_{\vert j_p}\,, \eeq where the subscript $|j$ stands
for a spatially covariant derivative.

\subsection{Iwasawa decomposition of the spatial metric}

 We systematically use the Iwasawa decomposition of the spatial
metric $g_{ij}$ and write \be \label{Iwasawaex} g_{ij} =
\sum_{a=1}^d e^{- 2 \b^a} {\cn^a}_i  \, {\cn^a}_j \ee where $\cn$
is an upper triangular matrix with $1$'s on the diagonal.
We will also need the Iwasawa coframe $\{ \theta^a \}$,
\be\label{Iwasawa1} \theta^a = {\cn^a}_i \, dx^i\,, \ee as well as
the vectorial frame $\{ e_a \}$ dual to the coframe $\{ \theta^a
\}$, \be\label{Iwasawa2} e_a = {\cn^i}_a \frac{\partial}{\partial
x^i} \ee where the matrix ${\cn^i}_a$ is the inverse of
${\cn^a}_i$, i.e., ${\cn^a}_i {\cn^i}_b = \delta^a_b$.  It is
again an upper triangular matrix with 1's on the diagonal. Let us
now examine how the Hamiltonian action gets transformed when one
performs, at each spatial point, the Iwasawa decomposition
\Ref{Iwasawaex} of the spatial metric. The kinetic terms of the
metric and of the dilaton in the Lagrangian (\ref{keyaction}) are
given by the quadratic form \be\label{dsigma2} G_{\mu\nu}d\beta^\mu d\beta^\nu =\sum_{a=1}^d (d
\beta^a)^2 - \left(\sum_{a=1}^d d \beta^a\right)^2  + d\phi^2 , Ê\quad \beta^\mu = (\beta^a,\phi).\ee The change of variables $(g_{ij}\to \beta^a, {\cn^a}_i )$ corresponds
to a point transformation and can be extended to the momenta as a canonical transformation in the standard way via \be\label{cantra}
\p^{ij}\dot{g}_{ij} \equiv \sum_a \pi_a \dot{\b}^a + \sum_{a}
{P^i}_a \dot{{\mathcal N}^a}_{i} \,\,. \ee Note that the momenta
\be\label{Nmomenta} {P^i}_a = \frac{\partial\mathcal L}{\partial
\dot{{\mathcal N}^a}_i} = \sum_{b} e^{2(\beta^b - \beta^a)}
{\dot\cn^a}_{\;\;j} {\cn^j}_b {\cn^i}_b \ee conjugate to the
nonconstant off-diagonal Iwasawa components ${\cn^a}_i$ are only
defined for $a<i$; hence the second sum in (\ref{cantra}) receives
only contributions from $a<i$.

\subsection{Splitting of the Hamiltonian}

We next split the Hamiltonian  density\, $\ch$  (\ref{Ham}) in two
parts:  ${\mathcal H}_0$, which is the kinetic term
for the local scale factors $\beta^\mu= (\beta^a, \phi)$, and $\cv$,
a ``potential density'' of
weight 2, which contains everything else. Our
analysis below will show why it makes sense to group the kinetic
terms of both the off-diagonal metric components and the $p$-forms
with the usual potential terms, i.e. the term $\mathcal M$ in
(\ref{Ham1}).  Thus, we write \be \ch =  {\mathcal H}_0 + \cv
\label{HplusV} \ee with the kinetic term of the $\b$ variables
\be\label{eq3.23} {\mathcal H}_0 = \frac{1}{4}\, G^{\mu\nu}
\pi_\mu \pi_\nu\,, \ee where $G^{\mu\nu}$ denotes the inverse of
the metric $G_{\mu\nu}$ of Eq.~(\ref{dsigma2}). In other words,
the right hand side of Eq.~(\ref{eq3.23}) is defined by
\begin{equation}\label{Gmunuup}
G^{\mu \nu} \pi_\mu \pi_\nu \equiv \sum_{a=1}^d \pi_a^2 -
\frac{1}{d-1} \left(\sum_{a=1}^d \pi_a\right)^2 + \pi_\phi^2\,,
\end{equation}
where $ \pi_\mu \equiv (\pi_a, \pi_\phi)$ are the momenta
conjugate to $\beta^a$ and $\phi$, respectively, i.e. \be \pi_\m =
2 \tilde{N}^{-1} G_{\m \n} \dot{\beta}^\n = 2 G_{\m \n} \frac { d
{\beta}^\n}{d\tau}\, . \ee The total (weight 2) potential density,
\be \cv = \cv_S + \cv_G + \sum_p \cv_{p}  + \cv_\phi\, , \ee is
naturally split into a ``centrifugal'' part $\cv_S$ linked to the kinetic
energy of the off-diagonal components (the index $S$ referring to
``symmetry,''), a ``gravitational'' part $\cv_G$, a term
from the $p$-forms, $\sum_p \cv_{p}$, which is a sum of an ``electric'' and a
``magnetic'' contribution and also a  contribution to
the potential coming from the spatial gradients of the dilaton
$\cv_\phi$.

\begin{itemize}

\item{``centrifugal'' potential}
\be \label{centrifugal} \cv_S = \frac{1}{2} \sum_{a<b}
e^{-2(\beta^b - \beta^a)} \left( {P^j}_b {{\mathcal
N}^a}_j\right)^2, \ee

\item{ ``gravitational'' (or ``curvature'') potential}

\be \label{gravitational} \cv_G =  - g R\, = \frac{1}{4}
{\sum_{a\neq b \neq c}} e^{-2\a_{abc}(\beta)} (C^a_{\; \; bc})^2 -
\sum_a e^{-2 \m_a(\beta)} F_a\,, \ee where  \be \a_{abc}(\beta)\equiv \sum_e \beta^e + \beta^a-\beta^b-\beta^c,\, a\neq b, b\neq c, c\neq a\ee and \be d\theta^a =-\frac{1}{2}C^a_{\; \; bc}\theta^b\wedge\theta^c\ee while $F_a$ is a polynomial of
degree two in the first derivatives $\partial \beta$ and of degree one
in the second derivatives $\partial^2 \beta$. 

\item{$p$-form potential}
\be
 \cv_{(p)} = \cv_{(p)}^{el} + \cv_{(p)}^{magn}\,,
\ee which is a sum of an ``electric'' $\cv_{(p)}^{el}$ and a
``magnetic'' $\cv_{(p)}^{magn}$ contribution. The ``electric''
contribution can be written as \beq \cv_{(p)}^{el} &=& \frac{e^{-
\lambda_p \phi}} {2 \,
p!} \, \pi_{(p)}^{j_1 \cdots j_p} \pi_{(p) \, j_1 \cdots j_p} \nonumber\\
&=& \frac{1}{2 \, p!} \sum_{a_1, a_2, \cdots, a_p} e^{-2 e_{a_1
\cdots a_p}(\b)} (\ce^{a_1  \cdots a_p})^2 \,,\eeq where $
\ce^{a_1 \cdots a_p} \equiv {\cn^{a_1}}_{j_1} {\cn^{a_2}}_{j_2}
\cdots {\cn^{a_p}}_{j_p} \pi^{j_1 \cdots j_p}\,,$  and $e_{a_1
\cdots a_p}(\b)$ are the "electric wall" forms, \be e_{a_1 \cdots
a_p}(\b) = \b^{a_1} + \cdots + \b^{a_p} + \frac{\l_p}{2} \phi  \,.
\ee And the ``magnetic'' contribution reads,

\beq \cv_{(p)}^{magn} &=&  \frac{e^{ \lambda_p \phi}}{2 \; (p+1)!}
\, g \, F^{(p)}_{j_1 \cdots j_{p+1}} F^{(p) \, j_1 \cdots j_{p+1}}
\nonumber\\&=&
\frac{1}{2 \, (p+1)!} \sum_{a_1, a_2, \cdots, a_{p+1}} e^{-2
m_{a_{1}  \cdots a_{p+1}}(\b)} (\cf_{a_1  \cdots a_{p+1}})^2 \,.
\eeq where $ \cf_{a_1 \cdots a_{p+1}} = {\cn^{j_1}}_{a_1} \cdots
{\cn^{j_{p+1}}}_{a_{p+1}} F_{j_1 \cdots j_{p+1}}\, $ and the
$m_{a_{1} \cdots a_{p+1}}(\b)$ are the magnetic linear forms \be
m_{a_{1} \cdots a_{p+1}}(\b) = \sum_{b \notin \{a_1,a_2,\cdots
a_{p+1}\}} \!\b^b - \frac{\l_p}{2}\, \phi\,, \ee
\item{dilaton potential}
\beq \cv_\phi  &=& g g^{ij}
\partial_i \phi \partial_j \phi\ \\
&=& \sum_a e^{-\mu_a(\beta)} (\mathcal{N}_a{}^i \partial_i
\phi)^2,. \eeq where \beq \mu_a(\b) = \sum_e \b^e - \b^a \eeq

\end{itemize}

\subsection{Appearance of sharp walls in the BKL limit}

In the decomposition of the hamiltonian as $\mathcal{H} = \mathcal{H}_0 +
\cv$, $\mathcal{H}_0$ is the kinetic term for the $\beta^\mu$'s while all other variables now only appear through the potential $\cv$ which is
schematically of the form \beq \label{V1} \cv(
 \b^\m, \partial_x \b^{\m}, &P&,Q)
=\sum_A c_A( \partial_x \b^{\m}, P,Q) \exp\big(-
2 w_A (\beta) \big)\,, \eeq  where $(P,Q) = ({\cn^a}_i, {P^i}_a,
\ce^{a_1 \cdots a_p},\cf_{a_1 \cdots a_{p+1}})$. Here $w_A (\beta)
= w_{A \m} \b^\m$ are the linear wall forms already introduced above: \beq \mbox{symmetry walls}&:& w^S_{ab}\equiv \beta^b - \beta^a; \quad a<b\nonumber\\ \mbox{gravitational walls}&:& \a_{abc}(\b) \equiv \sum_e \beta^e + \beta^a-\beta^b-\beta^c,\, a\neq b, b\neq c, c\neq a\nonumber\\ &\,& \m_a(\beta)\equiv\sum_e \beta^e -\beta^a,\nonumber\\ \mbox{electric walls}&:& e_{a_1 \cdots a_p}(\b)\equiv\beta^{a_1}+...+\beta^{a_p} + \frac{1}{2}\lambda_p\phi,\nonumber\\ \mbox{magnetic walls}&:& m_{a_1 \cdots a_{p+1}}(\b)\equiv\sum_e \beta^e -\beta^{a_1}-...-\beta^{a_{p+1}}-   \frac{1}{2}\lambda_p\phi. \nonumber \eeq  In order to take the limit $t\rightarrow 0$ which corresponds to
$\b^{\m} $ tending to future time-like infinity, we decompose $\b^{\m}$
into hyperbolic polar coordinates $(\rho,\g^{\m})$, i.e. \beq
\b^{\m} = \rho \g^{\m} \eeq where $\g^{\m}$ are coordinates on the
future sheet of the unit hyperboloid which are constrained by \beq
G_{\m\n} \g^{\m} \g^{\n} \equiv \g^{\m} \g_{\m} = -1\eeq and
$\rho$ is the time-like variable defined by \beq \rho^2 \equiv - G_{\m \n}
\b^{\m} \b^{\n} \equiv - \b_{\m} \b^{\m} >0, \eeq which behaves like $\rho\sim -\ln t \to +\infty$ at the BKL limit. In terms of these
variables, the potential term looks like

\beq \sum_A c_A(  \partial_x \b^{\m}, P,Q) \rho^2 \exp\big(- 2
\rho w_A (\g) \big)\,. \eeq
The essential point now is that, since $\r \to +
\infty$, each term $\rho^2 \exp\big(- 2 \rho w_A (\gamma) \big)$
becomes a {\it sharp wall potential}, i.e. a function of  $w_A
(\gamma)$ which is zero when $w_A (\gamma) >0$, and  $+\infty$
when $w_A (\gamma) < 0$. To formalize this behavior we define the
sharp wall $\Theta$-function\,\footnote{\,One should more properly
write $\Theta_\infty(x)$, but since this is the only step function
encountered here, we use the simpler notation
$\Theta(x)$.} as \be \Theta (x) := \left\{ \begin{array}{ll}
                      0  & \mbox{if $x<0$} \,,\\[1mm]
                      +\infty & \mbox{if $x>0$}\,.
                      \end{array}
                      \right.
\ee A basic formal property of this $\Theta$-function is its
invariance under multiplication by a positive quantity. Because
all the relevant prefactors $c_A( \partial_x \b^{\m}, P,Q)$ are
generically {\it positive} near each leading wall, we can formally
write
\begin{eqnarray}
\lim_{\rho\rightarrow\infty} && \Big[ c_A( \partial_x \b^{\m},Q,P) \rho^2
\exp\big(-\rho w_A (\gamma) \Big] = c_A(Q,P)\Theta\big(-2
w_A (\gamma) \big) \nonumber\\[2mm]
&\equiv &\Theta\big(- 2 w_A (\gamma) \big) \, 
\end{eqnarray}
valid in spite of the increasing of the spatial gradients \cite{Damour:2002et}.
Therefore, the limiting dynamics is equivalent to a free motion in
the $\b$-space interrupted by reflections against hyperplanes in
this $\b$-space given by $w_A (\b) = 0$ which correspond to a
potential described by infinitely high step functions

\beq \cv(\b, P,Q) = \sum_A \Theta\big(-2 w_A (\g) \big) \eeq The
other dynamical variables (all variables but the $\b^\mu$'s) completely disappear from this
limiting Hamiltonian and therefore they all get frozen as $t\rightarrow
0$.
\section{Cosmological singularities and Kac--Moody algebras}

Two kinds of motion are possible according to the volume of the
billiard table on which it takes place, i.e. the volume (after projection on hyperbolic space)
of the region where $\cv = 0$ for $t\to 0$, also characterized by the conditions,

\beq w_A(\b) > 0 \quad \forall A . \eeq
Depending on the fields present in the Lagrangian, on their dilaton-couplings and on the spacetime dimension, the
(projected) billiard volume is either finite or infinite. The
finite volume case corresponds to never-ending, chaotic oscillations for the $\beta$'s
while in the infinite volume
case, after a finite number of reflections off the walls, they tend to
an asymptotically monotonic Kasner-like behavior, see figure 3.

\begin{figure}[ht]
\centerline{\includegraphics[scale=.8]{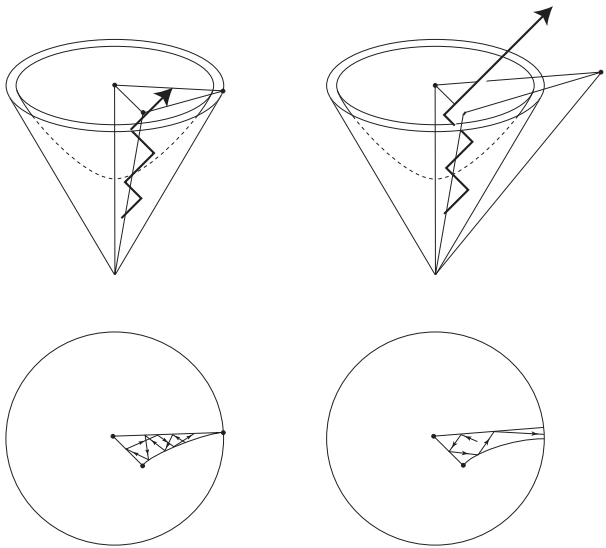}}
\caption{\small{Sketch of billiard tables describing the asymptotic cosmological behavior of Einstein-matter systems.}}
\end{figure}

In figure 3 the upper panels
are drawn in the Lorentzian space spanned by $(\beta^{\mu}) =
(\beta^a , \phi)$. The billiard tables are represented as
``wedges'' in $(d+1)$--dimensional (or  $d$--dimensional, if there are no dilatons)
 $\beta$-space, bounded by
hyperplanar walls $w_A (\beta) = 0$ on which the
billiard ball undergoes spe\-cular reflections. The upper left
panel is a  (critical) ``chaotic'' billiard table (contained
within the $\beta$-space future  light cone), while the upper
right one is a (subcritical) ``nonchaotic'' one (extending beyond
the light cone). The lower panels represent the corresponding
billiard tables (and billiard motions) after projection onto
hyperbolic space  $H_d$ ($H_{d-1}$ if there are no dilatons). The latter
projection is defined in the text by central projection onto
$\gamma$-space ({i.e.} the unit hyperboloid $G_{\mu\nu} \,
\gamma^{\mu} \, \gamma^{\nu} = -1$, see the upper panels), and is
represented in the lower panels by its image in the Poincar\'e ball (disk).


In fact, not all the walls are relevant for determining
the billiard table. Some of the walls stay behind the
others and are not met by the billiard ball. Only a subset of the walls $w_A(\b)$, called
dominant walls and here denoted $\{w_i(\b)\}$ are needed to delimit the hyperbolic domain. Once
the domiant walls are found, one can compute the following matrix
\beq A_{ij} \equiv 2 { w_i.w_j \over w_{i}.  w_{i} } \eeq where
$w_i . w_j = G^{\m\n} w_{i\m} w_{j\n }$. By definition,  the diagonal elements  are all equal to 2. Moreover, in many interesting cases, the off-diagonal elements happen to be non positive integers. These are precisely the characteristics of a generalized Cartan matrix, namely that  of an infinite KM algebra (see appendix). As
recalled in the introduction, for pure gravity in $D$ space-time dimensions, there are $D-1$ dominant walls and the
matrix $A_{ij}$ is exactly the generalized Cartan matrix of the hyperbolic KM
algebra $AE_{D-1} \equiv A^{\wedge\wedge}_{D-3} \equiv A^{++}_{D-3}$  which is hyperbolic for $D\leq10$. More generally, bosonic string theory in $D$
space-time dimensions is related to the Lorentzian
KM algebra $DE_D$ \cite{DH3,DdBHS} which is the canonical
Lorentzian extension of the finite-dimensional Lie algebra $D_{D-2}$.
The various superstring theories, in the critical dimension
$D=10$, and $M$-theory have been found \cite{DH3} to be related
either to $E_{10}$ (when there are two supersymmetries,
i.e. for type IIA, type IIB and $M$-theory) or to $BE_{10}$ (when
there is only one supersymmetry, i.e. for type I and II
heterotic theories), see the table.
\noindent The hyperbolic KM algebras are those relevant for chaotic billiards since their fundamental Weyl chamber has a finite volume. 

\vglue 3mm

\begin{centering}
\begin{tabular}{|c|p{6.5cm}|}
\hline Theory & Corresponding Hyperbolic KM algebra \\  \hline Pure gravity in
$D \leq 10$ & \scalebox{.5} {
\begin{picture}(180,60)
\put(5,-5){$\alpha_{1}$} \put(45,-5){$\alpha_2$}
 \put(125,-5){$\alpha_3$}  \put(50,45){$\alpha_{D-1}$}\put(85,-5){$\alpha_4$}
  \put(140,45){$\alpha_5$}
\thicklines \multiput(10,10)(40,0){4}{\circle{10}}
\multiput(15,10)(40,0){3}{\line(1,0){30}}
\multiput(90,50)(40,0){2}{\circle{10}}
\put(130,15){\line(0,1){30}} \put(50,15){\line(1,1){35}}
\dashline[0]{2}(95,50)(105,50)(115,50)(125,50)
\end{picture}
}
 \\ \hline
 M-theory, IIA and  IIB Strings & \scalebox{.5}{
\begin{picture}(180,60)
\put(5,-5){$\alpha_{1}$} \put(45,-5){$\alpha_2$}
\put(85,-5){$\alpha_3$}
 \put(125,-5){$\alpha_4$}
  \put(165,-5){$\alpha_5$} \put(205,-5){$\alpha_6$}
  \put(245,-5){$\alpha_7$}   \put(285,-5){$\alpha_8$}
  \put(325,-5){$\alpha_9$}
  \put(260,45){$\alpha_{10}$}
\thicklines \multiput(10,10)(40,0){9}{\circle{10}}
\multiput(15,10)(40,0){8}{\line(1,0){30}}
\put(250,50){\circle{10}} \put(250,15){\line(0,1){30}}
\end{picture}
 }
 \\ \hline
 type I and heterotic Strings & \scalebox{.5}{
\begin{picture}(180,60)
\put(5,-5){$\alpha_{1}$}
\put(45,-5){$\alpha_2$}\put(85,-5){$\alpha_3$}
 \put(125,-5){$\alpha_{4}$}
  \put(165,-5){$\alpha_{5}$}
\put(205,-5){$\alpha_{6}$} \put(245,-5){$\alpha_{7}$}
\put(285,-5){$\alpha_{8}$} \put(325,-5){$\alpha_{9}$}
  \put(70,45){$\alpha_{10}$}
\thicklines \multiput(10,10)(40,0){9}{\circle{10}}
\multiput(15,10)(40,0){7}{\line(1,0){30}}
\dashline[0]{2}(95,10)(105,10)(115,10)(125,10)
\put(295,7.5){\line(1,0){30}}\put(295,12.5){\line(1,0){30}}
\put(305,0){\line(1,1){10}} \put(305,20){\line(1,-1){10}}
\put(90,50){\circle{10}} \put(90,15){\line(0,1){30}}
\end{picture}
} \\ \hline
closed bosonic string in $D=10$ & \scalebox{.5}{
\begin{picture}(180,60)
\put(5,-5){$\alpha_{1}$} \put(45,-5){$\alpha_2$}
\put(85,-5){$\alpha_3$}
 \put(125,-5){$\alpha_4$}
  \put(165,-5){$\alpha_5$} \put(205,-5){$\alpha_6$}
  \put(245,-5){$\alpha_7$}   \put(285,-5){$\alpha_8$}
  \put(100,45){$\alpha_9$}
  \put(260,45){$\alpha_{10}$}
\thicklines \multiput(10,10)(40,0){8}{\circle{10}}
\multiput(15,10)(40,0){7}{\line(1,0){30}}
\put(250,50){\circle{10}} \put(250,15){\line(0,1){30}}
\put(90,50){\circle{10}} \put(90,15){\line(0,1){30}}
\end{picture} }\\ \hline
\end{tabular}
\end{centering} 

\vglue 3mm

\noindent {\small{This table displays the Coxeter--Dynkin diagrams which encode the geometry of the
billiard tables describing the asymptotic cosmological behavior of General Relativity and of
three blocks of string theories: ${\mathcal B}_2 = \{$$M$-theory,
type IIA and type IIB superstring theories$\}$, ${\mathcal B}_1 =
\{$type I and the two heterotic superstring theories$\}$, and
${\mathcal B}_0 = \{$closed bosonic string theory in $D=10\}$.
Each node of the diagrams represents a dominant wall of the
cosmological billiard. Each Coxeter diagram of a
billiard table corresponds to the Dynkin diagram of a (hyperbolic) KM algebra: $E_{10}$,
$BE_{10}$ and $DE_{10}$}} . 

\vglue 3mm
 The precise links between a chaotic billiard and its corresponding Kac--Moody
algebra can be summarized as follows
\begin{itemize}
\item the scale factors $\b^{\mu}$ parametrize a Cartan element $h = \sum_{\m=1}^{r} \b^{\m}h_{\mu}
$,
\item  the dominant walls $w_i(\b), (i=1,...,r)$ correspond to the simple roots
$\a_i$ of the KM algebra,
 \item the group of reflections in the
cosmological billiard is the Weyl group of the KM algebra,
and
\item the billiard table can be identified with the Weyl chamber
of the KM algebra.
\end{itemize}

\section{$E_{10} $ and a ``small tension" limit of SUGRA$_{11}$}

The main feature of the gravitational billiards that can be
associated with the KM algebras is that there exists a group
theoretical interpretation of the billiard motion: the asymptotic
BKL dynamics is equivalent (in a sense to be made precise below),
at each spatial point, to the asymptotic dynamics of a
one-dimensional nonlinear $\sigma$-model based on a certain
infinite-dimensional coset space $G/K$, where the KM group $G$ and
its maximal compact subgroup $K$ depend on the specific model. As
we have seen, the walls that determine the billiards are the {\it
dominant walls}. For the KM billiards, they correspond to the {\it
simple roots} of the KM algebra. As we discuss below, some of the
subdominant walls also have an algebraic interpretation in terms
of higher-height positive roots. This enables one to go beyond the
BKL limit and to see the beginnings of a possible identification
of the dynamics of the scale factors {\em and} of all the
remaining variables with that of a nonlinear $\sigma$-model
defined on the  cosets of the KM group divided by its
maximal compact subgroup \cite{DHN2,Damour:2002et}. 


For concreteness, we will only consider one specific example here:
the relation between the cosmological evolution of $D=11$
supergravity and a null geodesic on $E_{10} / K(E_{10})$
\cite{DHN2} where $KE_{10}$ is the maximally compact subgroup of  $E_{10}$. The $\sigma$-model is formulated in terms of a one-parameter dependent group element $\cv=\cv(t)\in E_{10}$ and its Lie algebra value derivative \be v(t) :=\frac{d\cv}{dt}\cv^{-1}(t) \in e_{10}.\ee The action is \be S_1^{E_{10}} = \int{dt \over n(t)} <v_{sym}(t)\vert v_{sym}(t)> \ee with a lapse function $n(t)$ whose variation gives rise to the Hamiltonian constraint ensuring that the trajectory is a null geodesic. The symmetric projection \be v_{sym} := \frac{1}{2} (v+v^T)\ee  is introduced in order to define an evolution on the coset space. 
Here $< . \vert . >$
is the standard invariant bilinear form on $E_{10}$ ; $v^T$ is the ``transpose"
of $v$ defined with the Chevalley involution\footnote{The Chevalley involution is defined by $\omega(h_i ) =-h_i ; \
\omega(e_i ) =-f_i ; \ \omega(f_i ) = -e_i $} 
as $v^T =- \omega(v)$. This action is
invariant under $E_{10}$,

\beq \cv(t) \rightarrow k(t) \cv(t) g \hspace{1cm} \mbox{where}\hspace{1cm} k \in KE_{10} \
g\in E_{10} \eeq Making use of the explicit Iwasawa parametrization of
the generic $E_{10}$ group element $\cv = K A N$ together with the gauge choice $K= 1$ (Borel gauge), one can write
$$ {{\mathcal V}}(t) = \exp X_h (t) \cdot \exp X_A (t)$$
with $X_h(t) = {h^a}_b  {K^b}_a$ and $$X_A (t) = \ft1{3!} A_{abc}
E^{abc} + \ft1{6!} A_{a_1\dots a_6} E^{a_1\dots a_6} + \ft1{9!}
A_{a_0|a_1 \dots a_8} E^{a_0|a_1 \dots a_8} + \dots\, .$$ Using
the $\E$ commutation relations in $GL(10)$ form together with the
bilinear form for $\E$, one obtains up to height 30 \footnote{We keep
only the generators $E^{abc}, \ E^{a_1\dots a_6}$ and $E^{a_0|a_1
\dots a_8}$ corresponding to the $E_{10}$ roots $\a = \sum n_i \a_i $ with height $\sum_i
n_i \leq 29 $ ($\a_i$ are simple roots and $n_i$ integers)} ,

\begin{eqnarray}\label{Lag}
n {{\mathcal L}} &=& \ft14 (g^{ac} g^{bd} - g^{ab} g^{cd}) \dot
g_{ab} \dot g_{cd}
  + \ft12 \ft1{3!} DA_{a_1a_2a_3}DA^{a_1a_2a_3} \non[2mm]
&&  
 + \ft12 \ft1{6!} DA_{a_1 \dots a_6}DA^{a_1\dots a_6}
  + \ft12 \ft1{9!} DA_{a_0 | a_1 \dots a_8} DA^{a_0|a_1\dots a_8}\,,
\end{eqnarray}
where $g^{ab} = {e^a}_c {e^b}_c$ with $$ {e^a}_b \equiv {(\exp
h)^a}_b\,,$$ and all ``contravariant indices'' have been raised by
$g^{ab}$. The ``covariant'' time derivatives are defined by (with
$\partial A\equiv \dot A$)
\begin{eqnarray}\label{Dtime}
DA_{a_1a_2a_3} &:=& \partial A_{a_1a_2 a_3} \,,\nonumber\\[2mm]
DA_{a_1\dots a_6} &:=& \partial A_{a_1 \dots a_6}
    + 10 A_{[a_1a_2 a_3} \partial A_{a_4a_5 a_6]} \,,\non[2mm]
DA_{a_1|a_2\dots a_9} &:=& \partial A_{a_1|a_2 \dots a_9}
    + 42 A_{\langle a_1a_2 a_3} \partial A_{a_4 \dots  a_9 \rangle} \non[2mm]
&& 
- 42 \partial A_{\langle a_1a_2 a_3} A_{a_4 \dots  a_9 \rangle}
    + 280 A_{\langle a_1a_2 a_3} A_{a_4a_5 a_6} \partial A_{a_7a_8
a_9\rangle} \, .~~~~~~~~~
\end{eqnarray}
Here antisymmetrization $[\dots]$, and projection on the $\ell =
3$ representation $\langle \dots \rangle$, are  normalized with
strength one (e.g. $[[\dots]] = [\dots]$). Modulo field
redefinitions, all numerical coefficients in \Ref{Lag} and in
\Ref{Dtime} are uniquely fixed by the structure of $\E$. 

In order to compare the above coset model results with those of the bosonic part of $D=11$ supergravity, we recall the action
\begin{eqnarray}
\label{sugra} S^{sugra}_{11} &= &\int d^{11} x \Bigl[ \sqrt{-{\rm G}} \, R({\rm
G}) - \frac{\sqrt{-{\rm G}}}{48} \, {\mathcal
F}_{\alpha\beta\gamma\delta} \, {\mathcal
F}^{\alpha\beta\gamma\delta} \nonumber \\
&&+ \frac{1}{(12)^4} \, \varepsilon^{\alpha_1 \ldots \alpha_{11}}
\, {\mathcal F}_{\alpha_1 \ldots \alpha_4} \, {\mathcal
F}_{\alpha_5 \ldots \alpha_8} \, {\mathcal A}_{\alpha_9
\alpha_{10} \alpha_{11}} \Bigl]\,.
\end{eqnarray}
The space-time indices $\alpha , \beta , \ldots $  take the values  $ 0,1,
\ldots , 10$;  $\varepsilon^{01 \ldots 10} = +1$, and 
the four-form ${\mathcal F}$ is the exterior derivative of
${\mathcal A}$, ${\mathcal F} = d {\mathcal A}$. Note the presence
of the Chern--Simons term ${\mathcal F} \wedge {\mathcal F} \wedge
{\mathcal A}$ in the action (\ref{sugra}). Introducing a
zero-shift slicing ($N^i=0$) of the eleven-dimensional space-time,
and a {\em time-independent} spatial zehnbein $\theta^a(x) \equiv
{E^a}_i(x) dx^i$, the metric and four-form ${{\mathcal F}} = d\cA$
become
\begin{eqnarray}\label{Gauge}
  &&  ds^2 = {\rm G}_{\alpha\beta} \, dx^{\alpha} \, dx^{\beta} = - N^2
(d{x^0})^2 + G_{ab} \theta^a \theta^b \\ \!\!\!\!\!\! {{\mathcal
F}} &=& \frac1{3!}{{\mathcal F}}_{0abc}\,  dx^0
\!\wedge\!\theta^a\!\wedge\! \theta^b\!\wedge\! \theta^c +
\frac1{4!}{{\mathcal F}}_{abcd} \,
\theta^a\!\wedge\!\theta^b\!\wedge\! \theta^c\!\wedge\!\theta^d .
\nn
\end{eqnarray}
We choose the time coordinate $x^0$ so that the lapse
$N=\sqrt{G}$, with $G:= \det G_{ab}$ (note that $x^0$ is not the
proper time\,\footnote{\,In this section, the proper time is denoted
by $T$ while the variable $t$ denotes the parameter of the
one-dimensional $\sigma$-model introduced above.} $T = \int N
dx^0$; rather, $x^0\rightarrow\infty$ as $T \rightarrow 0$). In
this frame the complete evolution equations of $D=11$ supergravity
read
\begin{eqnarray}\label{EOM}
\partial_0 \big( G^{ac} \partial_0 G_{cb} \big)  &=&
\ft16 G {{\mathcal F}}^{a\beta\gamma\delta} {{\mathcal
F}}_{b\beta\gamma\delta} - \ft1{72} G {{\mathcal
F}}^{\alpha\beta\gamma\delta} {{\mathcal
F}}_{\alpha\beta\gamma\delta} \delta^a_b 
- 2 G {R^a}_b (\Gamma,C)\,, \nonumber\\[2mm]
\partial_0 \big( G{{\mathcal F}}^{0abc}\big) &=&
\ft1{144} \varepsilon^{abc a_1 a_2 a_3 b_1 b_2 b_3 b_4}
          {{\mathcal F}}_{0a_1 a_2 a_3} {{\mathcal F}}_{b_1 b_2 b_3 b_4} \nonumber\\[1mm]
  &&  
 + \ft32 G {{\mathcal F}}^{de[ab} {C^{c]}}_{de} - G {C^e}_{de} {{\mathcal F}}^{dabc}
     - \partial_d \big( G{{\mathcal F}}^{dabc} \big)\,, \nonumber\\[2mm]
\partial_0 {{\mathcal F}}_{abcd} &=& 6 {{\mathcal F}}_{0e[ab} {C^e}_{cd]} + 4
\partial_{[a} {{\mathcal F}}_{0bcd]}\,,
\end{eqnarray}
where $a,b \in \{1,\dots,10\}$ and $\alpha,\beta \in
\{0,1,\dots,10\}$, and $R_{ab}(\Gamma,C)$ denotes the spatial
Ricci tensor; the (frame) connection components are given by $ 2
G_{ad} {\Gamma^d}_{bc} = C_{abc} + C_{bca} - C_{cab} +
          \partial_b G_{ca} + \partial_c G_{ab} - \partial_a G_{bc}
$ with ${C^a}_{bc} \equiv G^{ad} C_{dbc}$ being the structure
coefficients of the zehnbein $d\theta^a = \frac12 {C^a}_{bc}
\theta^b \!\wedge\! \theta^c$. (Note the change in sign convention
here compared to above.) The frame derivative is $\partial_a
\equiv {E^i}_a (x) \partial_i$ (with $ {E^a}_i {E^i}_b =
\delta^a_b$). To determine the solution at any {\it given} spatial
point $x$ requires knowledge of an infinite tower of spatial
gradients; one should thus augment \Ref{EOM} by evolution
equations for $\partial_a G_{bc}, \partial_a {{\mathcal
F}}_{0bcd}, \partial_a {{\mathcal F}}_{bcde}$, etc., which in turn
would involve higher and higher spatial gradients.

The main result of concern here is the following: there exists a
{\it map} between geometrical quantities constructed at a given
spatial point $x$ from the supergravity fields $G_{\mu\nu}(x^0,x)$
and $\cA_{\mu\nu\rho}(x^0,x)$ and the one-parameter-dependent
quantities $g_{ab}(t), A_{abc} (t), \dots$ entering the coset
Lagrangian \Ref{Lag}, under which the supergravity equations of
motion \Ref{EOM} become {\it equivalent, up to 30th order in
height}, to the Euler-Lagrange equations of \Ref{Lag}. In the
gauge \Ref{Gauge} this map is defined by $t = x^0  \equiv \int dT/
\sqrt{G}$ and
\begin{eqnarray}\label{map}
g_{ab}(t) &=& G_{ab} (t,x) \,,\nonumber\\[3mm]
DA_{a_1a_2a_3}(t)  &=& {{\mathcal F}}_{0a_1 a_2 a_3} (t,x)\,,
\non[3mm] DA^{a_1 \dots a_6} (t) &=&   - \ft1{4!}
\varepsilon^{a_1\dots a_6 b_1 b_2 b_3 b_4} {{\mathcal F}}_{b_1 b_2
b_3 b_4} (t,x) \,, \non[3mm] DA^{b|a_1 \dots a_8 } (t)&=& \ft32
\varepsilon^{a_1\dots a_8 b_1 b_2} \big( {C^b}_{b_1 b_2} (x) +
\ft29 \delta^b_{[b_1}  {C^c}_{b_2] c} (x) \big)\,.
\end{eqnarray}

Let us also mention in passing (from \cite{DN04})
that the $\E$ coset action is not compatible with the addition of an
eleven-dimensional cosmological constant in the supergravity action
(an addition which has been proven to be incompatible with supersymmetry in
\cite{BDHS}).

\section{Conclusions}
We have reviewed the finding that
the general solution of many physically relevant (bosonic) Einstein-matter
systems, in the vicinity of a space-like
singularity, exhibits a remarkable mixture of chaos and symmetry. Near the singularity, the behavior
of the general solution is  describable, at each (generic)
spatial point, as a billiard motion in an auxiliary Lorentzian space or, 
after a suitable ``radial''  projection, as a
billiard motion on hyperbolic space. This motion appears to be chaotic in many physically
interesting cases including pure Einstein gravity in
any space-time dimension $D\leq 10$ and the particular Einstein-matter systems
arising in string theory. Also, for
these cases, the billiard tables can be identified with the Weyl chambers of
some Lorentzian Kac--Moody algebras. In the case of the bosonic sector of
supergravity in 11-dimensional space-time the underlying Lorentzian algebra
is that of the hyperbolic Kac--Moody group $E_{10}$, and there exists some
evidence of a correspondence between the general solution of the
Einstein-three-form system and a null geodesic in the infinite-dimensional
coset space $E_{10} / K (E_{10})$, where $K (E_{10})$ is the maximal
compact subgroup of $E_{10}$.

\vglue 8mm
\noindent{\bf Acknowledgement}

\noindent It is a pleasure to thank Sophie de Buyl and Christiane Schomblond for
their help in trimming the manuscript and in improving the figures.

\appendix
\section{Kac-Moody algebras}

A KM algebra $\mathcal{G}(A)$ can be constructed out of a
generalized Cartan matrix $A$, (i.e. an $r\times r$ matrix such that $A_{ii} = 2, i=1
,...,r$, ii) $-A_{ij} \in \mathbb{N}$ for $i\neq j$
and iii)  $A_{ij} = 0$ implies $A_{ji} = 0$) according to the following rules for the
Chevalley generators $\{ h_i, e_i, f_i \}, {i=1,...,r}$: 
\begin{eqnarray}
\ [ e_i ,f_j ] &=& \d_{ij} h_i \nonumber \\
\ [h_i ,e_j] &=& A_{ij} e_j \nonumber \\
\ [h_i ,f_j] &=& -A_{ij} f_j \nonumber \\
\ [h_i ,h_j] &=& 0. \nonumber
\end{eqnarray}
The generators must also obey the Serre's relations, namely
\begin{eqnarray}
(\mathrm{ad}\,e_i)^{1-A_{ij}} e_j &=& 0 \nonumber \\
(\mathrm{ad}\,f_i)^{1-A_{ij}} f_j &=& 0 \nonumber
\end{eqnarray} and the Jacobi identity. $\mathcal{G}(A)$ admits a triangular decomposition
\be \mathcal{G}(A) = n_- \oplus h \oplus n_+   \ee where
$n_-$ is generated by the multicommutators of the form
$[f_{i_1},[f_{i_2},...]]$, $n_+$ by the multicommutators of the
form  $[e_{i_1},[e_{i_2},...]]$ and $h$ is the Cartan
subalgebra. 

The algebras $\mathcal{G(A)}$
build on a symmetrizable Cartan matrix $A$ have been classified according to properties of their eigenvalues
\begin{itemize}
\item if $A$ is positive definite, $\mathcal{G(A)}$ is a finite
dimensional Lie algebra;
\item if $A$ admits one null eigenvalue and the others are all stricly positive, $\mathcal{G(A)}$ is an Affine KM algebra;
\item if $A$ admits one negative eigenvalue and all the others
are strictly positive, $\mathcal{G(A)}$ is a Lorentzian KM algebra.
\end{itemize}
A KM algebra such that the deletion of one node from its
Dynkin diagram gives a sum of finite or affine algebras is called an
\textit{hyperbolic} KM algebra. These algebras are all known; in particular,  there exists  no hyperbolic algebra with rang higher than 10. 

\end{document}